# Applications, challenges and ethical issues of AI and ChatGPT in education

1st Dimitrios Sidiropoulos
University of Aegean, Social Science School
Cultural Technology and Communication Dpt
Mytilene, Greece
dsidir@gmail.com

2nd Christos-Nikolaos Anagnostopoulos
University of Aegean, Social Science School
Cultural Technology and Communication Dpt
Mytilene, Greece
canag@ct.aegean.gr

*Abstract-* **Artificial Intelligence (AI) in recent years has shown an unprecedentedly impressive development, tending to play a catalytic role in all aspects of life. The interest of the academic community, but also of governments, is huge in the dynamics of AI and is reflected by the truly explosive amount of investment and research that is underway. Enthusiastic opinions and statements about AI are made every day, but at the same time they also bring to the fore alarming predictions about its effects. This paper aims to describe the opportunities emerging from the use of artificial intelligence and ChatGPT to improve education, but also to identify the challenges and ethical issues that arise.**

## I. INTRODUCTION

The rapid technological developments that have taken place in recent decades play a decisive role in various areas of human life and affect almost every aspect of daily life [1], [2]. Of course, education could not remain untouched by this technological revolution [3].

The evolution of the educational process over time has been dynamically transformed. The traditional model of learning as we know it is now considered obsolete, and teaching has undergone sweeping changes both in its structure and in the way information and knowledge are transmitted, creating a new landscape on the educational map. The addition of technology, interactive media, distance education, new approaches to implement personalized learning and recently Artificial Intelligence (AI), tend to shape a new reality in education [4].

The application of AI in education began to make its appearance dynamically in recent years, signaling a new revolution in the educational process [5] with the ultimate goal of the smooth integration of new proposals and technologies into the educational process. This new technology (AI) can theoretically offer a multitude of tools and benefits in the field of education, but first of all it should be tested in many ways while taking care of the concerns, doubts and ethical barriers of the educational community, resulting from its application.

The uniqueness of the nature of people, the diversity of characters but also the diversity of cognitive level, experiences and the way and pace of learning, demonstrate the need to adapt the educational process in a way that promotes and applies personalized learning techniques. Over the years, various algorithms, methods, and techniques have been proposed by researchers in order to integrate personalized learning capabilities into the educational process. However, given the complexity of human nature, the possible factors influencing how a learner learns can be so numerous that they are difficult to capture. However, certain factors have been proven to play a very important role in the way a student learns. Various machine learning algorithms, having a wealth of data at their disposal, can offer personalized learning options, creating almost unique learning paths for each learner. A catalytic role at this point promises to be offered by AI, which is expected to be an important addition, taking over part of the work of a teacher [6]. AI can create customized training materials, assessment sheets, provide feedback, be a personalized assistant - Intelligence Tutoring System [7], but also contribute to the evaluation of the learner.

Understandably, all of these new developments, especially since the advent of ChatGPT, have upset much of the educational community. Many trainers feel they will lose their jobs and AI will replace their role. Educators are also aware that they will likely face new risks related to privacy, plagiarism and data security. The role of the teacher can be transformed, to be the supervisor of the whole process, who will use AI in support [7], but in any case he should determine the goals and intended results of the educational process.

It is very important and should be an absolute priority of researchers as well as governments, the in-depth investigation of the possibilities, disadvantages and ethical barriers resulting from the integration of AI in various areas of everyday life and especially education. The educational community needs to act quickly, wisely and efficiently in order to smoothly absorb the positives that AI can offer in the education sector, while ensuring the quality and reliability of the education provided. The educational staff should join forces with AI, and governments should draw up a modern educational policy, in order to upgrade the education provided for the benefit of the learners.

All actors in the field of education should act collectively, in order to obtain the greatest possible benefits from the use of artificial intelligence in education, but at the same time to be properly protected from the risks it entails. It is imperative to properly use AI in education, to realize and exploit the key opportunities, but also to prevent or mitigate the extraordinary risks, as well as to deal with the unintended consequences in a timely manner.

## II. DESCRIPTION OF CHATGPT

ChatGPT, developed by Open Artificial Intelligence (OpenAI), is an advanced conversational AI, a large language model (LLM) that uses Deep Learning and allows humans to interact with a computer in a more natural way with the dialogue form [8], [9]. GPT, short for "Generative Pre-trained Transformer," refers to a series of natural language processing models developed by OpenAI, and was made available to the general public on November 30, 2022 [10]. At its core, ChatGPT uses machine learning techniques to generate human-like text based on the input prompts it receives. It is also known as a form of genetic artificial intelligence due to its capacity to generate unique outcomes. ChatGPT employs natural language processing to train with data from the web, providing users with answers to questions or prompts. It can understand the context of a conversation, answer questions, provide explanations and even produce creative content such as stories or poems. Generative corresponds to the model's ability to generate text rather than simply understand or classify it. Pretrained indicates that the model has undergone an initial training phase where it was learned from a large amount of Internet information. Transformer refers to the type of model architecture used, which helps understand the input text.

Models of this kind are trained on large text data sets including books, articles, web pages, so that they can predict the next word in a sentence and generate complete sentences in response to a question [11], [12]. In the case of ChatGPT, 570 GB of data were required, representing 300 billion words, while approximately 175 billion parameters have been provided to the system. This combination of technologies allows ChatGPT to generate meaningful responses that closely mimic natural human conversation [8], [13], [14], [15].

ChatGPT is a "large language model" (LLM), an artificial intelligence system or tool endowed with the ability to read, summarize, solve mathematical problems, produce learning materials, design syllabi, guide students , explain terms and abstract concepts and theories, find and correct errors in the source code of computer programs, write various types of essays, create advisory reports, identify scientific debates in various fields of study through access to relevant literature, translate texts into different languages, create a bibliographic list, and create complete sentences comparable to those of humans [13], [16].

ChatGPT utilizes machine learning, which presently stands as the leading method in the field of Artificial Intelligence (AI). Because of its ability to generate and evaluate information, ChatGPT can play an important role in teaching and learning processes. In addition to various types of AI, ChatGPT has the potential to enhance the educational process and experience for learners. It can function as an independent tool or be incorporated into different systems and platforms. [14].

Educators can utilize ChatGPT to offer tailored educational assistance to their students. Based on the individual requirements and learning preferences of each student, ChatGPT is capable of suggesting specialized learning materials and activities. For example, an educator can employ ChatGPT to evaluate student performance data and pinpoint areas where learners are having difficulties with particular concepts. Teachers can take the help of ChatGPT to answer students' questions and provide feedback to them. In addition, teachers can request clarifications and examples from ChatGPT on a specific topic to further increase the effectiveness of their teaching. Educators can also harness the capabilities of ChatGPT to evaluate student assignments and quizzes. ChatGPT can be used to check submitted work for plagiarism. It is also very interesting that the model can generate questions/quizzes based on different levels of difficulty (eg, high, medium, easy) for a variety of topics [17]. ChatGPT has the potential to produce impressive results for writing stories, poetry, songs, essays and other subjects, but it also has some limitations. Users can ask the Chatbot questions and it will answer persuasively and with arguments. ChatGPT's proficiency in grasping context and offering insightful analysis renders it an effective instrument for collecting, assessing, and comprehending market trends. This technology can enhance existing procedures and gather qualitative data via informal surveys, analyze data and extract features from massive amounts of unstructured data, and save researchers time and effort. ChatGPT differs from previous AI models in that it can write software in multiple languages, debug code, break down a complex topic into manageable chunks, prepare candidates for interviews, and write reports [10].

According to research by Firat M. [18], AI and ChatGPT can improve the performance of learners in different learning environments and also motivate them in this direction. The learning experience in educational environments can be enhanced using ChatGPT and increase learner engagement in e-courses by offering personalized suggestions for study according to the learning goals and needs of each learner. At the same time, self-learning is effectively promoted through personalized support, easy access to resources, real-time feedback and guidance, enhanced use of open educational resources, and self-assessment.

OpenAI has achieved considerable advancements in enhancing deep learning through the introduction of GPT-4. The release of GPT-4 in March 2023 has already generated widespread discussion, as the tool appears to be more accurate, reliable and adaptable than its predecessor [19]. This new multifunctional language model has the ability to accept not only text but also images and produce relevant results. It may not currently be as capable as a human in real-world scenarios, but it has performed very well in many professional and academic metrics. It clearly outperforms ChatGPT, which is promising for future releases [13]. GPT-4 is OpenAI's most advanced model, promising safer, more accurate and more useful answers. According to the company, GPT-4 can solve difficult problems more accurately thanks to its broader general knowledge and problem-solving abilities.

## III. APPLICATIONS OF CHATGPT

AI and ChatGPT can find application in many everyday tasks and in many aspects of human life. One of the most obvious application of ChatGPT in business is customer service, offering 24-hour support as a digital assistant and minimizing the requirement for human involvement. Users can engage with ChatGPT for tasks like organizing meetings, setting reminders, catching up on news, managing smart home devices, or sending messages, essentially through a new user-friendly communication experience. AI and ChatGPT can also be used as a mental health consultant, offering a wide range of support services, but also as a rich source of medical information for the provision of medical help and care. The multi-faceted role of ChatGPT allows it to act as a virtual legal assistant to offer legal advice [20].

For education in particular, the advantages of AI applications and ChatGPT can be varied and truly impressive. Enhancing learners' motivation for autonomous learning [21], automated assessment and grading of exercises and assignments [22], feedback and specific suggestions for improvement, virtual assistants – Intelligence Tutoring Systems (ITS), the possibility of personalized learning with adapted content, learning pace and teaching approach, the prevention of cheating and plagiarism as well as the preparation of adapted teaching materials and assessment sheets, are just some of the positives that we have to attribute to AI [3], [23]. ChatGPT's ability to understand human language, to be able to converse with a human, its flexibility to respond to various situations, its speed of response, its relatively low cost of use, its ability to be used as a 24/7 personal training assistant, are some more of its strong points [8]. As a study assistant, ChatGPT can support learners beyond conventional textbooks to understand difficult topics by providing additional explanations and step-by-step advice [20].

AI appears to have the potential to make a positive contribution to education, increasing personalized learning for all learners [24]. Intelligent instructional support systems are currently the most promising field of artificial intelligence for the transformation of education, and are perhaps the most effective tools for personalizing instruction [22]. Several training platforms have integrated ChatGPT as a virtual instructor in order to adapt the learning pace and preferences of the learners making the learning experience more engaging and interesting [20]. The growth of this personalization is currently in full development as researchers experiment with new learning models and as a result new opportunities are created in the educational field [8].

Using ChatGPT, instructors can develop unique assessment sheets and instructional content [19]. Instructors can use AI to reduce their workload, gain insights into learners, and promote innovation in the classroom. AI systems are designed to assist instructors by automating assessment, plagiarism detection, classroom management, and providing feedback [8]. Also, instructors can benefit from creating lesson plans, activities and exercises, providing tailored support and answers to learners' questions, immediate assessment and feedback [17].

A major advantage of ChatGPT is that it allows learners to learn through experimentation and experiences. ChatGPT can aid students in enhancing their reading and writing abilities by offering recommendations, such as for syntax and grammar. Additionally, the model is capable of creating practice exercises and quizzes across diverse subjects, including mathematics, physics, and language arts. It can also generate explanations and step-by-step solutions to a given problem and help develop problem-solving and analytical thinking skills [17]. Using ChatGPT, learners can evaluate different strategies and approaches to solving problems and achieving goals through play. It is therefore imperative for educators to turn challenges into opportunities [8].

Storytelling and creative writing suit ChatGPT, which as a creative partner can promote brainstorming. ChatGPT's ability to understand and correctly render content and suggestions makes it a vital resource for content creators. The model excels at writing reports [25], writing emails, creating marketing materials, and even managing social media. Also, ChatGPT as a virtual translator has shown promising results in various languages. ChatGPT is also proving to be a significantly promising tool for code generation in programming as well as a debugging aid in software development [17], [20].

ChatGPT offers both students and educators the chance to obtain precise responses to theoretical queries, particularly in the fields of communication, business writing, and music courses. The answers provided by ChatGPT are usually accurate. Learners, if they want instant answers on a topic, can consider ChatGPT as a reliable option. In contrast to search engines that deliver countless results which may sometimes lack precision and relevance, ChatGPT furnishes responses tailored to the user-defined criteria, within the confines of the answer's word limit. These responses can provide users with sufficient information, eliminating the need to sift through an extensive array of sources to assess their accuracy and reliability. ChatGPT also has the ability to provide students with customized answers for case study analysis, writing business correspondence and reports. These responses can provide learners with information on how to approach the case study and create appropriate relevant reports [26].

Each learner has a unique cognitive profile based on their own prior knowledge of a subject, their personal social background, their own financial, aesthetic and emotional situation. Teaching is most effective when it adapts to these changing contexts. AI can help identify learning gaps in each learner, offer tailored recommendations and content as well as step-by-step solutions to complex problems [23].

Several studies have shown that one of the functions pursued by the use of AI is the automated assessment of learners. [22]. Assessment in the educational context refers to any control or judgment of a task or a learner's performance. Assessment has been identified as one of the three pillars of school education along with curriculum, learning and teaching. The purpose of contemporary assessment is to measure what students know, what they have understood and what they can apply. Optimally, evaluations should encompass the entire spectrum of learners' skills and offer valuable insights into learning achievements. However, every student is unique and so are their learning paths. How standardized assessment can be used to assess each student, with unique abilities, skills, and preferences, is a question that requires much investigation [22].

## IV. CHALLENGES OF CHATGPT

Despite the advantages that AI and ChatGPT may have, there are limitations, challenges and disadvantages associated with its use that we should consider. Some of them are found in the unreliability of the information that produces when it does not have sufficient data and references resulting in providing wrong answers [8], it cannot use idioms intelligently, it is not able to correctly and accurately cite the sources or to evaluate their quality, there is an absence of ethical criteria, it shows occasional logical errors and grammatical errors, there is a possible bias in technology, it has a weakness in calculating complex and difficult mathematical expressions, and it ignores its lack of knowledge [16], [27].

ChatGPT has demonstrated remarkable capabilities regarding text production, however it has presented cases with incorrect or misleading information [8]. Guaranteeing the precision and reliability of information generated by AI is of immense importance in order to maintain the integrity of any work or research [20]. ChatGPT has been trained on a huge amount of data, which may contain bias and affect the actual results. ChatGPT has the potential to produce detrimental material, including hate speech or misinformation. Implementing safeguards to avert the creation of such content is crucial. Additionally, ChatGPT might exhibit biases towards specific cultural and linguistic groups, potentially resulting in skewed or unsuitable responses. Ensuring reliable results requires constant control, feedback and updating [24].

Although ChatGPT possesses a broad awareness and comprehension of many subjects, its depth of knowledge might be insufficient for certain specialized niche topics. A differential performance is also observed in different subject areas, including finance, programming, mathematics and general public questions [19]. Lack of quality and diversity in training data can lead to biased results with negative consequences. The decline in critical thinking, independence in problem solving, and skill development seem quite close to the increasing reliance on AI [24].

ChatGPT has access to the data of a huge number of users, which raises privacy and data protection concerns. Developing policies and regulations is crucial to guarantee the protection and responsible use of user data. Also, models like ChatGPT are quite complex and require huge computing resources and perhaps we should also consider the environmental footprint they leave in terms of energy consumption [24].

ChatGPT has the ability to exhibit critical thinking abilities and produce highly authentic text with limited data input, posing a potential risk to the credibility of online examinations. This is especially pertinent in higher education environments where such testing is increasingly prevalent. It is important for both educators and institutions to be aware of the possibility of using ChatGPT for cheating and to explore countermeasures in order to maintain the fairness and validity of online testing for all learners [12].

In general, ChatGPT has several limitations and disadvantages, including: i) Inaccurate or Misleading Information ii) Sensitivity to prompts iii) Overuse of certain phrases iv) Inability to control sources or access information in real time v) Difficulty handling ambiguous queries vi) Lack of contextual association vii) Ethically ambiguous content viii) Long chat texts ix) Inability to create visual content x) Addressing requests that are unsuitable or damaging xi) Challenges in identifying and adjusting to the user's level of expertise xii) Restricted capacity for emotional understanding xiii) Absence of individualized feedback xiv) Constrained expertise in specific fields xv) Incapacity to engage with external systems xvi) Difficulty in processing queries in multiple languages xvii) Struggles with interpreting figurative language xviii) Restricted creative abilities xix) Tendency to make overly broad generalizations xx) Variability in output quality xxi) Significant energy use and environmental impact during training xxii) Challenges in comprehending human intuition xxiii) Absence of self-awareness xxiv) Significant resources needed for training and development [8], [19], [20], [24], [28], [29].

## V. ETHICAL ISSUES OF CHATGPT

The influence of ChatGPT on education was swift and polarizing. Despite its wide-ranging potential in educational settings, numerous universities have prohibited its use because of concerns about plagiarism, and several countries have also implemented temporary bans. The main concerns according to a UNESCO report are found in the following: academic integrity, lack of use regulations, privacy concerns, cognitive bias, gender and diversity, accessibility, commercialization [7], [14].

The primary issue raised regarding ChatGPT in higher education centers on academic honesty. Universities and instructors are raising concerns about the heightened potential for plagiarism and cheating if students utilize ChatGPT for completing exams or assignments. [9]. Online exams have become commonplace in higher education. Given that ChatGPT can produce text resembling human writing, even in academic contexts, educators and academic institutions need to be vigilant about the potential for cheating in online exams through the use of ChatGPT. In short, ChatGPT threatens the fairness, transparency and validity of online examinations [17]. Concerns also exist regarding the potential ineffectiveness of current plagiarism detection tools when faced with text generated by ChatGPT. This situation has prompted the creation of new applications designed to identify whether artificial intelligence has been employed in writing assignments. Numerous universities globally have prohibited

ChatGPT over worries regarding academic honesty, while others have altered their methods of conducting assessments.

ChatGPT is not currently subject to any AI ethics regulation or control. ChatGPT's extremely rapid growth has alarmed many, prompting a coalition of scholars and industry leaders to issue an open letter advocating for a temporary halt in the advancement of AI systems. This outage will potentially allow time to prevent risks that need to be further investigated and understood so that appropriate protection protocols can be developed.

April 2023, Italy was the first nation to restrict ChatGPT, citing privacy issues. The Italian data protection agency stated that collecting and storing personal data for ChatGPT's training lacked a legal foundation. Additionally, ethical issues were pointed out due to the tool's incapacity to ascertain a user's age, potentially exposing minors to unsuitable content. This case underscores wider concerns about the collection, ownership, and usage of data by AI technologies.

It's crucial to understand that ChatGPT does not operate on ethical guidelines and is unable to discern between right and wrong or differentiate between truth and falsehood. ChatGPT acquires information exclusively from the databases and textual content it analyzes on the internet, thereby also absorbing any cognitive biases present in that data. Consequently, it's essential to scrutinize its outputs critically and cross-reference them with other information sources. These biases might manifest in the model's outcomes, leading to unequal treatment or the reinforcement of stereotypes. [20].

As the use of artificial intelligence in data processing and analysis grows, it's natural that worries about the privacy and security of personal data are also escalating. Ensuring the protection of sensitive personal data is of the utmost importance. Increasingly, there are also concerns about intellectual property issues, as AI models contribute to the writing of research ideas, hypotheses, and even entire texts and papers. ChatGPT, similar to other machine learning models, may exhibit biases if it is trained on data that contains biases. These biases may lead to unfair results for specific individuals or communities, especially in sectors like employment, healthcare, and the justice system. Other times it can lead to misinformation, harmful advice or even malicious content. Consequently, the model might internalize these biases and generate replies that are derogatory, thereby reinforcing detrimental stereotypes. In addition, it is important to ensure the security and privacy of sensitive data, since ChatGPT may access and process such data. Standards such as data anonymization, encryption, and privacy must be incorporated to protect users' personal data and prohibit breaches. ChatGPT could be exploited for nefarious activities, including disseminating false information, creating fabricated news and content, or participating in online harassment - cyberbullying. ChatGPT may be susceptible to scenarios where harmful actors deliberately generate false information, compelling the model to yield undesirable or damaging outcomes. The use of ChatGPT in swaying human actions and choices brings up issues regarding personal autonomy. The processing power needed to train ChatGPT and execute its commands can lead to notable environmental effects, such as high energy use and carbon emissions. [20], [24].

Overall, ChatGPT, similar to other AI models, is prone to a range of biases, encompassing those related to gender, race, culture, language, and ideology. These biases originate from the training data of the model, which is composed of content generated by humans and sourced from the internet. ChatGPT in general has many biases: (i) biases related to gender, race, and culture, (ii) bias in language, (iii) bias in ideology, (iv) stereotype confirmation bias, diversity, (v) time bias, (vi) exclusion bias of specific groups, communities and negative feelings towards them, (vii) bias in favor of commercial interests, (ix) bias in perception and judgment, (x) bias towards specific points of focus, (xi) bias due to the way information is presented, (xii) bias originating from particular sources, (xiii) bias towards new or novel information, (xiv) bias towards either positive or negative sentiments, (xv) bias at extreme levels, (xvi) unconscious or subtle bias, (xvii) bias based on fundamental principles or beliefs, (xviii) bias towards more recent information, (xix) bias influenced by collective thinking or consensus. (xx) information availability bias, (xxi) false consent bias, (xxii) possibility overestimation bias [20], [24], [28], [29].

## VI. CONCLUSION

It is clear from the evidence presented in this research that AI technology has reached extraordinary levels and is now capable not only of retrieving and providing simple information, but also of critical thinking to an extent. Among ChatGPT's most remarkable capabilities is its aptitude for critical thinking, coupled with its skill in articulating thoughts and ideas in flawlessly written language. It has demonstrated exceptional ability and results in various fields, to the point of simulating the abilities of humans [12].

While ChatGPT has made significant strides in creating comprehensive responses, there is room for improvement in handling extended conversations. Prompting methods and techniques can allow users to actively contribute to managing ChatGPT responses (Gill et al., 2023). A promising future direction is to add multiple inputs as well as outputs, such as audio and images. Future research should focus on bias prevention and equity enhancement to reduce bias. Research should provide advanced solutions for maintaining privacy, confidentiality, and personal data protection [20].

The implementation of AI and ChatGPT in education also requires a number of immediate changes in order for this transition to be smooth and with minimal negative effects. Teachers should improve the structure of assignments and exams to students, using more interactive tools in order to reduce the possibility of plagiarism. Instructors should be informed and trained in order to detect plagiarism, but also to incorporate AI into their teaching preparation. It is of utmost importance to educate students about the disadvantages of ChatGPT, including its reliance on faulty data, lack of current knowledge, and the potential to produce false or misleading findings [19].